\documentstyle[amsfonts,amsthm,amssymb,amscd,amstext,epsfig]{article}
\input{epsf}

\newcommand{\be}{\begin{equation}}
\newcommand{\ee}{\end{equation}}
\newcommand{\bea}{\begin{eqnarray}}
\newcommand{\eea}{\end{eqnarray}}
\newcommand{\ba}{\begin{array}}
\newcommand{\ea}{\end{array}}

\def\bbox{{\,\lower0.9pt\vbox{\hrule \hbox{\vrule height 0.2 cm\hskip
0.2 cm \vrule height 0.2 cm}\hrule}\,}}
\newcommand{\dsl}{\pa \kern-0.5em /}

\newcommand{\nn}{\nonumber \\}

\def\ben{\begin{equation}}
\def\een{\end{equation}}
\def\bena{\begin{eqnarray}}
\def\eena{\end{eqnarray}}
\def\ft#1#2{{\textstyle{{\scriptstyle #1}\over {\scriptstyle #2}}}}

\def\e{\epsilon}
\def\6{\partial}

\def\G{\Gamma}
\def\a{\alpha}
\def\g{\gamma}

\def\d{\delta}

\def\q{\theta}
\def\p{\phi}
\def\pa{\partial}

\def\today{\ifcase\month\or  January\or February\or March\or April\or
May\or June\or  July\or August\or September\or October\or November\or
December\fi \space\number\day, \number\year}

\def\bZ {\mathbb{Z}}
\def\bR {\mathbb{R}}

\def\kk{\bar{k}}
\def\bbbone{{\mathchoice {\rm 1\mskip-4mu l} {\rm 1\mskip-4mu l}
{\rm 1\mskip-4.5mu l} {\rm 1\mskip-5mu l}}}

\begin{document}

\begin{flushright}
DAMTP-2004-46\\
CECS-PHY-04-10\\
hep-th/0405214
\end{flushright}

\begin{center}

\vspace{2cm}

{\LARGE {\bf Deforming baryons into confining strings} }

\vspace{1cm}

{\large Sean A. Hartnoll $^\sharp $ and Rub\'en Portugues $^\ddagger$

\vspace{0.9cm}

{\it ${}^\sharp$ DAMTP, Centre for Mathematical Sciences, Cambridge University \\
 Wilberforce Road, Cambridge CB3 OWA, UK \\}

\vspace{0.3cm}

e-mail: s.a.hartnoll@damtp.cam.ac.uk

\vspace{0.5cm}

{\it ${}^\ddagger$ Centro de Estudios Cientificos (CECS)\\
Avenida Arturo Prat 514, Casilla 1469, Valdivia, Chile }

\vspace{0.3cm}

e-mail: rp@cecs.cl }

\end{center}

\vspace{1cm}

\begin{abstract}

We find explicit probe D3-brane solutions in the
infrared of the Maldacena-Nu\~nez background. The solutions describe
deformed baryon vertices: $q$ external quarks are separated in
spacetime from the remaining $N-q$. As the separation is taken to
infinity we recover known solutions describing infinite confining strings
in ${\mathcal{N}}=1$ gauge theory. We present
results for the mass of finite confining strings as a function of
length. We also find probe D2-brane solutions in a confining type IIA
geometry, the reduction of a $G_2$ holonomy M theory background.
The relation between these deformed baryons and confining strings
is not as straightforward.

\end{abstract}

\pagebreak
\setcounter{page}{1}

\tableofcontents

\setcounter{equation}{0}

\vspace{.5truecm}

\section{Introduction}

The advent of the gravity/gauge theory correspondence
\cite{Maldacena:1997re,Gubser:1998bc,Witten:1998qj} has opened a
new window into strongly coupled field theories. Consequently, a
lot of effort has been invested in finding supergravity duals of
${\mathcal N}=1$ theories in four dimensions. Two of the most studied
backgrounds are \cite{Klebanov:2000hb,Maldacena:2000yy}. Such
solutions allow the use of supergravity analysis to study strong
coupling properties of the dual QCD-like field theories.

Within the dual backgrounds, probe brane analysis has been used
extensively to understand the gravity counterparts of
nonperturbative field theory objects such as baryons, mesons and
confining strings. The initial studies considered brane probes in
$AdS_5\times S^5$, the dual geometry to ${\mathcal N} =4$
Yang-Mills theory. Following
\cite{Witten:1998xy,Gross:1998gk}, D5-brane probes wrapping the
$S^5$ were used to construct explicitly the baryon vertex
\cite{Callan:1998iq,Imamura:1998gk,Gomis:1999xs}. The wrapped
brane was identified as the baryon vertex because the background
has $N$ units of Ramond-Ramond (RR) five-form flux on the $S^5$:
through the Wess-Zumino term of the probe D5-brane action the flux
acts as a source for $N$ fundamental strings on the worldvolume.
The strings then join the vertex to external quarks. The work of
\cite{Callan:1998iq} was extended in \cite{Callan:1999zf} to
find numerically solutions describing baryons in which a fraction
of the $N$ quarks were pulled apart from the rest in spacetime.
When the quarks are far apart, the branes were found to describe the
confining strings of the gauge theory. The probe brane description of
confining strings was further considered in \cite{Herzog:2001fq},
using results from \cite{Pawelczyk:2000ah,Pawelczyk:2000hy,Bachas:2000ik}.

In this work we find explicit analytic solutions of the full
second order Dirac-Born-Infeld (DBI) equations of motion. These
will be nonsupersymmetric probe branes in the infrared of supergravity
geometries which are dual to ${\mathcal N}=1$ confining gauge
theories. The solutions describe baryon vertices where $q$ quarks are
being pulled apart from the remaining $N-q$. In the limit of infinite
separation, the solutions recover known infinite confining string
solutions \cite{Herzog:2001fq,Callan:1999zf}. The fact that our
solutions are very explicit will enable us to calculate analytically
various properties of the deformed baryons. For instance, we
calculate the mass of finite confining strings as a function of
length.

A schematic illustration of our solutions is given in Figure 1 in
section 2.2 below. Figure 2 shows a couple of solutions more
explicitly. The
solutions are one dimensional from a spacetime perspective. At each
point the remaining probe brane directions partially wrap an
internal sphere of the background.

We focus on two background geometries in particular: the
Maldacena-Nu\~nez solution \cite{Maldacena:2000yy} of IIB
supergravity and a IIA solution which results from dimensional
reduction of M theory on a manifold of $G_2$
holonomy
\cite{Cvetic:2001ih,Cvetic:2001kp,Brandhuber:2001kq,Edelstein:2002zy}.
These backgrounds both describe the near horizon geometry of wrapped
branes; some comparisons are made in \cite{Gursoy:2003hf}.

The baryon vertex of ${\mathcal N}=1$ Yang-Mills theory in these
backgrounds is given by a wrapped probe brane in an entirely
analogous way to the $AdS_5\times S^5$ case. In the
Maldacena-Nu\~nez solution the baryon vertex is a D3-brane
wrapping a nontrivial $S^3$ of the background \cite{Loewy:2001pq}.
For the IIA background it has been proposed that
the baryon vertex is a D2-brane wrapping an $S^2$ \cite{Acharya:2001dz}.

Amongst the deformed baryons we consider, in the Maldacena-Nu\~nez
case it is easy to identify the finite length confining strings. In the
IIA background, there does not seem to be a limit of the deformed
baryons that is immediately connected with confining strings.

\section{IIB background: The Maldacena-Nu\~nez solution}

\subsection{The infrared background: ${\mathcal{M}}^4\times S^3$}

The Maldacena-Nu\~nez background \cite{Maldacena:2000yy} is a solution of type IIB string
theory describing the result of the geometric
transition induced by D5-branes wrapping an $S^2$ in the resolved
conifold. The background preserves 4 supercharges and is dual to ${\mathcal{N}}=1$
super Yang Mills theory, modulo issues of decoupling of Kaluza-Klein
and little string theory modes.

The far infrared of the field theory is described by the $r\rightarrow
0$ region of the background. The background collapses at $r=0$ to
${\mathcal{M}}^4\times S^3$ with $N$ units of RR flux through the sphere
\bea
\label{smallr}
ds^2_{\text{IIB}} &=& e^{\Phi_{0}}\left[ dx_{1,3}^2 + \alpha' N \left(d\psi^2
  + \sin^2 \psi [d\q^2 + \sin^2 \q \, d\p^2 ] \right)
  \right] \nn
C_2^{RR} &=& - \alpha' N \left( \psi - \frac{1}{2} \sin 2\psi
\right) \sin \q \, d\q \wedge d\p \,,
\eea
where $\Phi_0$ is the value of the dilaton at the origin.
The RR field strength is thus
\be
F_3^{RR} = d C_2^{RR} = - 2 \alpha' N \sin^2\psi \sin\q d\psi \wedge d\q
  \wedge d\p = - 2 \alpha' N {\mbox{vol}}_{S^3} \,.
\ee
The ranges
of the angles are $0 \leq \psi \leq \pi$, $0 \leq \q \leq \pi$, $0
\leq \p < 2\pi$. We are working in the string frame.

The Dirac-Born-Infeld action for a probe D3-brane with these
background fluxes is
\be
S_{\text{DBI}} = \int d^4\xi {\mathcal{L}} = - T_3 \int d^4\xi  \,
e^{-\Phi}
\sqrt{ -\det (^\star G +
  {\mathcal F} ) } + T_3 \int {\mathcal F} \wedge ^\star C_2^{RR} \,,
\ee
where as usual $T_p = 1/[(2\pi)^p \alpha'^{(p+1)/2}] $ and
here ${\mathcal F} = 2\pi\alpha' F$. We use $^\star G$ and
$^\star C_2^{RR}$ to denote the pullback onto the worldvolume of
the metric and the RR potential.

We will find solutions to the DBI equations of motion in the
background at $r=0$, given in (\ref{smallr}). Setting $r(\xi)=0$
is a consistent truncation of the full equations of motion.

\subsection{Probe D3-brane solutions}

We are looking for solutions describing deformed baryons. To
this end,
we look for probe D3-brane solutions with the following ansatz. The
ansatz can be thought of as describing fundamental strings extended
in the $x$ direction that the RR flux has blown up at each point, in
an Emparan-Myers effect \cite{Emparan:1997rt,Myers:1999ps}, to a
D3-brane `wrapping' an $S^2$ in the $S^3$ of the background. See
Figures 1 and 2 below.
\bea\label{eq:d3embedding}
t & = & \a'^{1/2} \xi_0 \,, \qquad x = \a'^{1/2} \xi_1 \,, \qquad
\psi = \psi(\xi_1) \,, \qquad \q =
\xi_2 \,, \qquad \p = \xi_3 \,, \nonumber \\
A & = & k(\xi_1) \xi_0 d\xi_1 \qquad \Rightarrow \qquad F =
k(\xi_1) d\xi_0 \wedge d\xi_1 \,.
\eea
Plugging this ansatz into the full DBI equations of motion, one
finds that the equations are solved if the functions $k(\xi_1)$
and $\psi(\xi_1)$ satisfy the following relations
\bea\label{MNsolutions}
\kk^2 & = & \left(\psi - \sin\psi\cos\psi \right)^2 \,, \nonumber \\
\left[\frac{d\psi}{d\xi_1} \right]^2 & = & \frac{e^{-2\Phi_0}}{N}
\left[-e^{2\Phi_0}+C^4\sin^4\psi + C^4\left(\kk + \frac{2\pi
    k_0}{C^2} \right)^2 \right] \,,
\eea
where we have introduced
\be
\kk = \frac{2\pi}{C^2} (k-k_0) \,.
\ee
Note that the solution has two dimensionless constants $C,k_0$.
The range of $k_0$ when positive is restricted to
$[e^{\Phi_0}/2\pi,\infty)$. For negative $k_0$ the allowed range is
more complicated and will be discussed below. One may take either the
positive or negative square roots in (\ref{MNsolutions}) to obtain the
same solutions.

It is perhaps
surprising that we can obtain such an explicit form for the
solutions. The solutions are given precisely, up to an integral.
We have found a two parameter family of solutions to the full
nonlinear second order DBI equations of motion in the infrared
background. We have not restricted to supersymmetric solutions.
Indeed in the Appendix we show that none of our solutions are
supersymmetric. The first equation in (\ref{MNsolutions}) has
appeared before in similar configurations
\cite{Camino:1999xx,Camino:2001at} and
is essentially Gauss's law for the electric field.

The equations (\ref{MNsolutions}) have solutions where $\psi$ goes
from $0$ to $\pi$. The spatial sections of these solutions are
topologically $S^3$, with the worldvolume $S^3$ in the
nontrivial homology class of the background $H_3({\mathcal{M}}^4\times
S^3)$. However, the solutions are not just sitting at a point in
${\mathcal{M}}^4$ wrapping the $S^3$, but have an extension along the
$x$ direction given by
\be\label{eq:IIBlength}
\Delta\, x = \a'^{1/2}\int_0^\pi \frac{d\xi_1}{d\psi} d\psi \,.
\ee
Two limits at fixed $C$ illustrate the possible behaviours:
\bea\label{MNlength}
\Delta\, x & \sim & \left(\frac{\a' N e^{2\Phi_0}}{4}\right)^{1/2}
\frac{1}{k_0} + \cdots \,, \qquad \text{as} \qquad k_0 \to \infty \,, \nonumber
\\
\Delta\, x & \sim & 2.8044 \left[\frac{\a' N e^{\Phi_0}}{4}\right]^{1/2} \left[\frac{9\pi}{8 C^4}
\right]^{1/6} \frac{1}{\left(2 \pi k_0 - e^{\Phi_0}
  \right)^{1/6}} + \cdots\,, \quad \text{as} \quad k_0 \to \frac{e^{\Phi_0}}{2
  \pi} \,,
\eea
where the numerical factor comes from an elliptic integral.
Therefore the limits correspond to short and large extensions in
spacetime. Note that in the large length limit only the $\psi=0$
end goes to infinity, due to a pole in $d\xi_1/d\psi$ from
(\ref{MNsolutions}) at $\psi=0$. The $\psi=\pi$ end remains at a finite
position. We can obtain a pole at $\psi=\pi$ rather than $\psi=0$ by
noting that the transformation $\{ \psi \to \pi - \psi, k_0 \to
-C^2/2-k_0\}$ leaves the 
solutions (\ref{MNsolutions}) invariant. This symmetry will appear
later as the symmetry $q \leftrightarrow N-q$.
We will see below that it is a different
large length limit that is related to confining strings.

The induced spatial metric of the D3-branes is given by
\be
ds^2_{\text{D3}} = \a' N e^{\Phi}
\left[\left(\left[\frac{d\psi}{d\xi_1}\right]^2 + \frac{1}{N}\right)
  d\xi_1^2 + \sin^2\psi \left(d\xi_2^2 + \sin^2\xi_2 d\xi_3^2 \right) \right] \,.
\ee
As $\psi \to 0,\pi$, the induced geometry has a conical
singularity. Far from being a problem, this is exactly what one should
expect. The spatial sections of the D3-branes are compact, yet the
Wess-Zumino term in the DBI action is a source for the abelian
gauge field on the brane. The total charge on the compact sections
must be zero, so there must be more sources. These will be fundamental
strings attached at the conical points.

We will now understand the conical singularities as being due to
attached fundamental strings. The endpoints of fundamental strings
are electric sources for the worldvolume gauge field on the D3-branes
\cite{Callan:1997kz,Gibbons:1997xz}. The equations of motion in
the presence of external sources at $\psi=0$ and $\psi=\pi$ become
\be
\pa_i \frac{\pa {\mathcal{L}}}{\pa \pa_i A_j} = \frac{\pa
  {\mathcal{L}}}{\pa A_j} + Q^j_{0}\d^{(3)}(x+\Delta x) + Q^j_{\pi} \d^{(3)}(x)  \,.
\ee
In this expression, we should integrate the Wess-Zumino term in the
action by parts,
so that it contributes to the equations of motion as part of the
$\frac{\pa{\mathcal{L}}}{\pa A_j}$ term.
The electric charges may be read off the solutions as
\bea
Q^0_{0} & = & \lim_{\psi \to 0}
\int_{\xi_2=0}^{\pi} \int_{\xi_3=0}^{2\pi} \frac{\pa
  {\mathcal{L}}}{\pa \pa_1 A_{0}} d\xi_2 d\xi_3 \,, \nonumber \\
Q^0_{\pi} & = & - \lim_{\psi \to \pi}
\int_{\xi_2=0}^{\pi} \int_{\xi_3=0}^{2\pi} \frac{\pa
  {\mathcal{L}}}{\pa \pa_1 A_{0}} d\xi_2 d\xi_3 \,.
\eea
This is just Gauss's law on the brane worldvolume.
Applying the formulae to our solutions
\be
Q^0_0 = -\frac{2 N k_0}{C^2} \,, \qquad \qquad Q^0_{\pi} = N +
\frac{2 N k_0}{C^2} \,.
\ee
The total charge due to sources at the conical points is
$N$. One can check that this
precisely cancels the contribution from the Wess-Zumino term. Let
us write $Q^0_{\pi} = N - q$ and $Q^0_0 = q$. In the full
string theory $q \in \bZ$, corresponding to $q$ string endpoints.

The interpretation of the solutions now becomes clear. The baryon
vertex with $N$ external quarks is being pulled apart along the
$x$ axis with $N-q$ quarks at one end and $q$ at the other.
The configuration is illustrated in Figure 1 below.
\begin{figure}[h]
\begin{center}
\epsfig{file=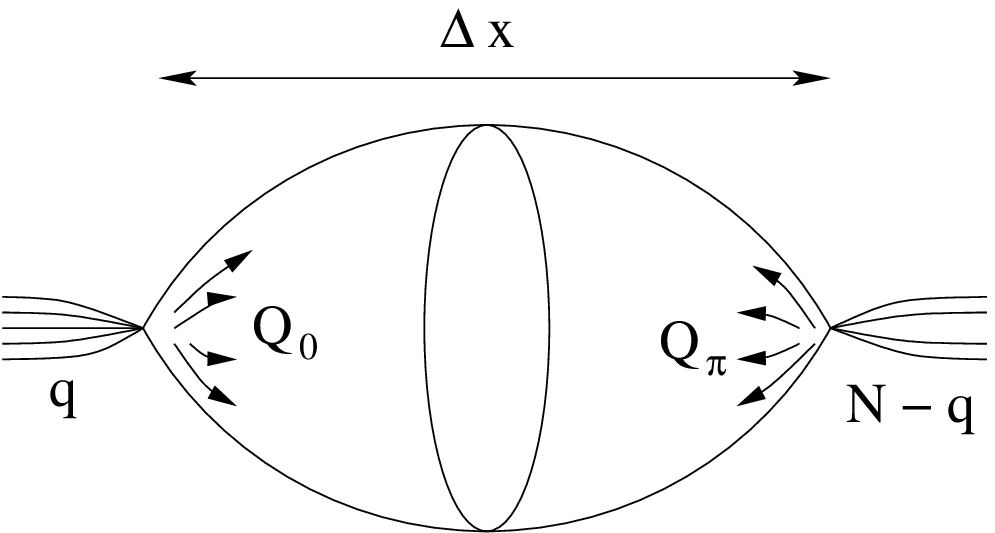,width=6cm}
\end{center}
\noindent {\bf Figure 1:} D3-brane pulled apart by $N-q$ and $q$
fundamental strings.
\end{figure}

The next step is to understand the energetics of the deformation
process. The energy density of the solution is given by
\be
{\mathcal{E}} = \frac{\pa {\mathcal{L}}}{\pa \dot{X}^\mu} \dot{X}^\mu
+\frac{\pa {\mathcal{L}}}{\pa \dot{A}} \cdot \dot{A} -
{\mathcal{L}} \,,
\ee
where the $X^\mu$ denote the spatial coordinates of the background and
$\dot{}$ denotes differentiation with respect to $\xi_0$.
Considered as an object in four dimensional spacetime, the tension is
\be\label{eq:tens}
T(\xi_1) = \int d\xi_2 d\xi_3 {\mathcal{E}} = \frac{N \left[C^4
\sin^2\psi - \left(C^2\psi + 2\pi
  k_0\right) \left(2C^2\sin\psi\cos\psi - C^2\psi-2\pi k_0 \right) \right]}{\a'^{1/2} 2\pi^2 C^2}
\ee
The mass of the solution is $M = \int d\xi_1 T(\xi_1)$. We can
calculate the masses in the two limits considered previously in (\ref{MNlength})
\bea\label{eq:IIBmasses}
M & \sim & \frac{N^2 e^{2\Phi_0}}{2C^2}
\frac{1}{\Delta x} + \cdots \,, \qquad \text{as} \qquad \Delta x
\to 0 \quad [k_0 \to
  \infty] \,, \nonumber \\
M & \sim & \frac{e^{\Phi_0}}{2\pi\a'} |q| \Delta x + \cdots \,,
\qquad \text{as} \qquad \Delta x \to \infty \quad \left[k_0 \to \frac{e^{\Phi_0}}{2
  \pi}\right] \,. \label{eq:collapse}
\eea
The short length limit takes us outside the regime of validity of
the DBI effective action, because $\pa_1 F \sim \a'^{1/2}/\Delta
x$ becomes large and $\a'$ corrections are important.
Therefore the mass divergence for short
extension should probably not be taken seriously, although it is
consistent with a $1/\Delta x$ Coulomb potential. The large separation
solution has energy linear in separation, as expected
for a confining theory. The large length expression has a simple
interpretation. It is the mass of $q$ fundamental strings in the
background (\ref{smallr})! This result is as we should expect in this
limit, where almost all the mass is coming from near the $\psi=0$
endpoint. Most of the D3-brane has collapsed to form fundamental
strings \cite{Emparan:1997rt}.

There is a different large separation limit that has a more interesting
physical interpretation. An infinite length occurs whenever $d\xi_1 /
d\psi$ has a pole at some $\psi_0 \in [0,\pi]$. In the large
length limit we have just considered, the pole is at the endpoint
$\psi_0 = 0$ when $k_0 = e^{\Phi_0}/2\pi$. From equation (\ref{MNsolutions})
one can see that there is another possibility. This is the interior
point $\psi_0 = -2\pi k_0 /C^2 = \pi q/N$ when $C^2
\sin\psi_0 = e^{\Phi_0}$. In this case one finds the mass
\be\label{eq:conf}
M \sim \frac{e^{\Phi_0} N}{2\pi^2 \alpha'} \sin \frac{\pi q
}{N} \, \Delta x + \cdots , \qquad \text{as}\qquad \Delta x \to
\infty
\quad \left[k_0 \to \frac{-C^2}{2\pi}\sin^{-1}\frac{e^{\Phi_0}}{C^2}\right]
\,.
\ee
Note that in this limit $k_0$ is negative and hence $q$ is
positive. 
The expression (\ref{eq:conf}) is immediately recognised as the sine
formula for the  mass of the confining strings of the theory
\cite{Hanany:1997hr,Douglas:1995nw,Herzog:2001fq}. The mass per unit length of the $q$th
confining string in the infinite length limit becomes
\be\label{eq:conften}
T = \frac{e^{\Phi_0} N}{2 \pi^2 \alpha'^{1/2}} \sin \frac{\pi q}{N} \,,
\ee
recovering the result obtained in
\cite{Herzog:2001fq}. Our solutions place the confining strings in a
larger context. The confining strings arise as an infinite length limit of a two
parameter family of explicit probe brane solutions describing
deformed baryon vertices.

\enlargethispage{0.5cm}

The difference between the two different long length limits,
(\ref{eq:collapse}) and (\ref{eq:conf}), is illustrated in Figure
2. We see how in the former case the brane collapses at one end, as we
noted following the tension calculation (\ref{eq:collapse}).
\begin{figure}[h]
\begin{center}
\epsfig{file=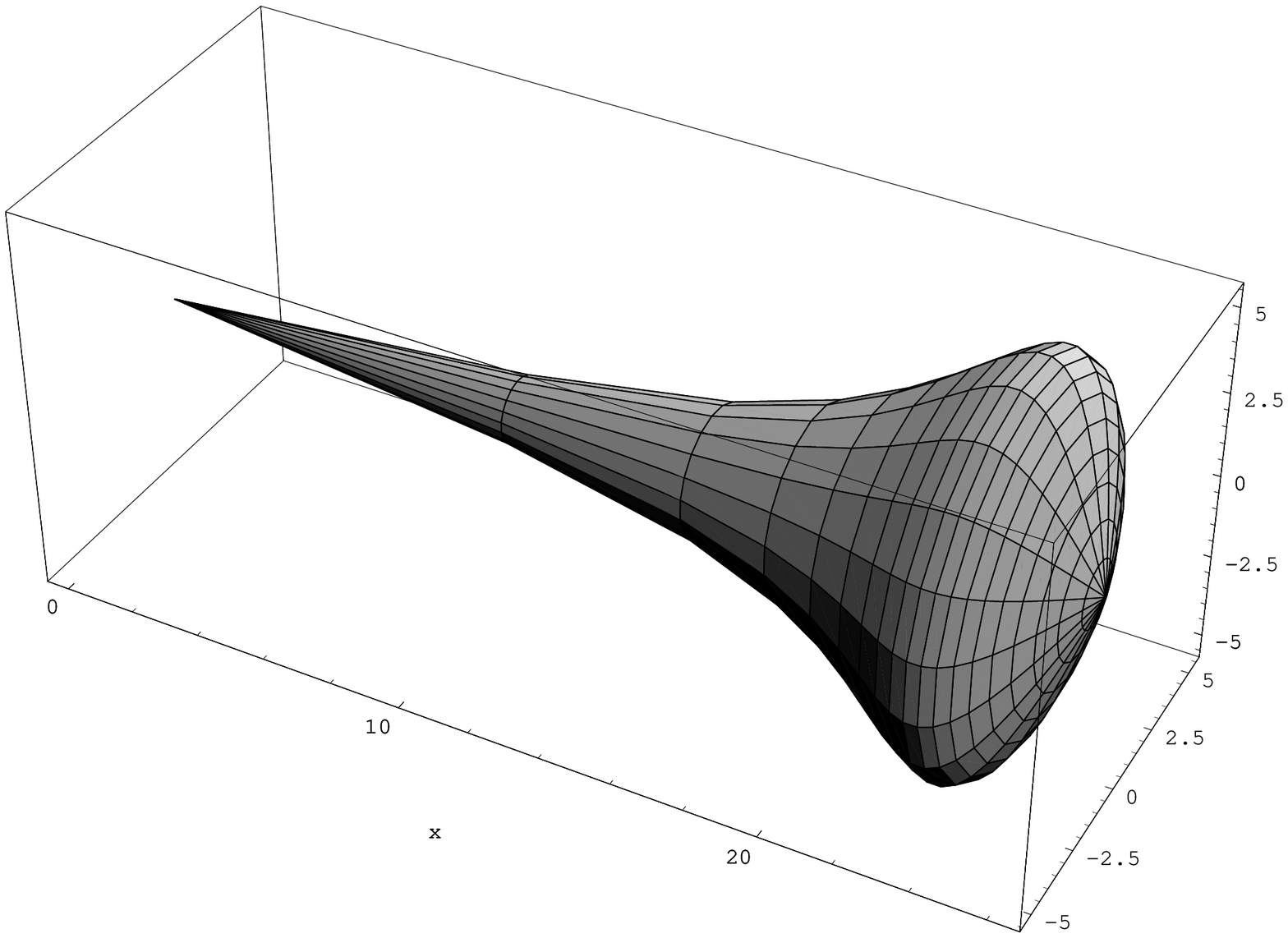,width=5cm}\hspace{1cm} \epsfig{file=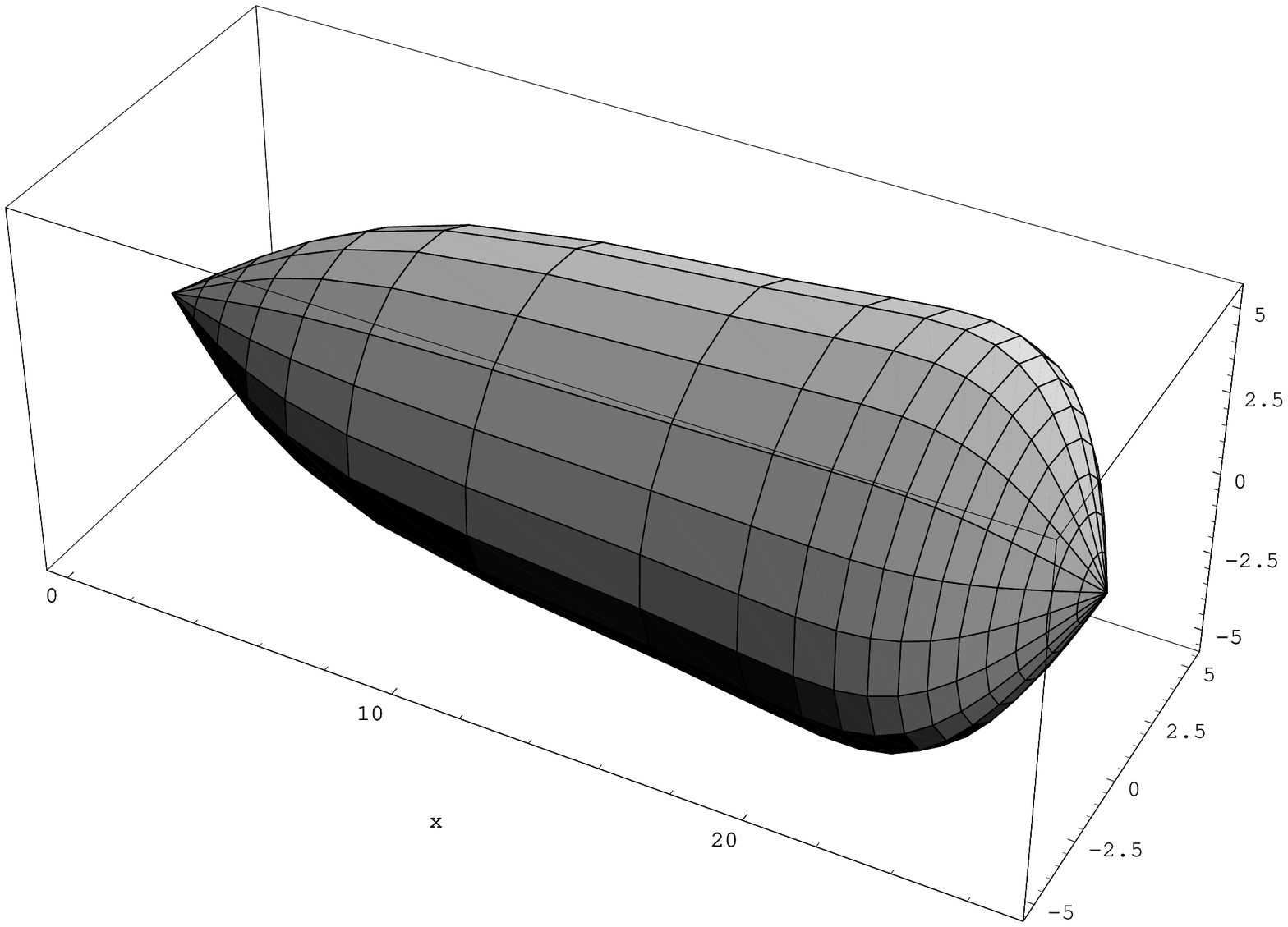,width=5cm}
\end{center}
\noindent {\bf Figure 2:} Probe D3-brane solutions with a large but
  finite extension in the $x$ direction. The background has $N=30$ and
  $\Phi_0=1$. The left hand configuration has $q=-10$ whilst the right
  hand configuration has $q=10$. One angular direction has been
  supressed.
\end{figure}

The physical reason for the difference between the two
cases is as follows. In the confining string case $q \leq N$ is
positive and hence the charges at each end, $Q_0^0$ and $Q_\pi^0$, are both
positive. The brane expands in a dielectric effect
\cite{Emparan:1997rt,Myers:1999ps} due to the RR flux on the $S^3$ of
the background, with slightly more expansion at the end with more charge.
However, in the collapsing case $q$ is negative. Therefore
the charge $Q_0^0$ is negative whilst $Q_\pi^0$ is positive.
The result is that the brane expands near the end with positive
charge but the negatively charged end collapses into fundamental strings.

The allowed range of negative $k_0$ with $C^2$ fixed is complicated
and we see in the next section that it is more natural to parameterise
in terms of $C$ with $q$ fixed. Suffice to note that if
$e^{\Phi_0} < C^2$ then one allowed range is
$k_0 \in (-C^2/2-e^{\Phi_0}/(2\pi),-C^2\sin^{-1}(e^{\Phi_0}/C^2)/(2\pi)]$. In
this case $-C^2\sin^{-1}(e^{\Phi_0}/C^2)/(2\pi) \leq -
e^{\Phi_0}/(2\pi)$ so the pole at $k_0=-e^{\Phi_0}/(2\pi)$ is not
reached.

\subsection{Energetics of finite length confining strings}

We can use the explicit expressions for the mass
(\ref{eq:tens}) and length (\ref{eq:IIBlength}) to examine the
energetics of the full space of probe D3-brane solutions. We would
like to calculate the mass of the D3-brane with a fixed length in
spacetime, $\Delta x$, and a fixed charge imbalance $q$. There is a
unique solution associated with a pair $(\Delta x,q)$, providing a more
physical parameterisation of the space of solutions than $(C,k_0)$.

Figure 3 is a plot of mass against length for finite length confining
strings at four values of $q$.
\begin{figure}[h]
\begin{center}
\epsfig{file=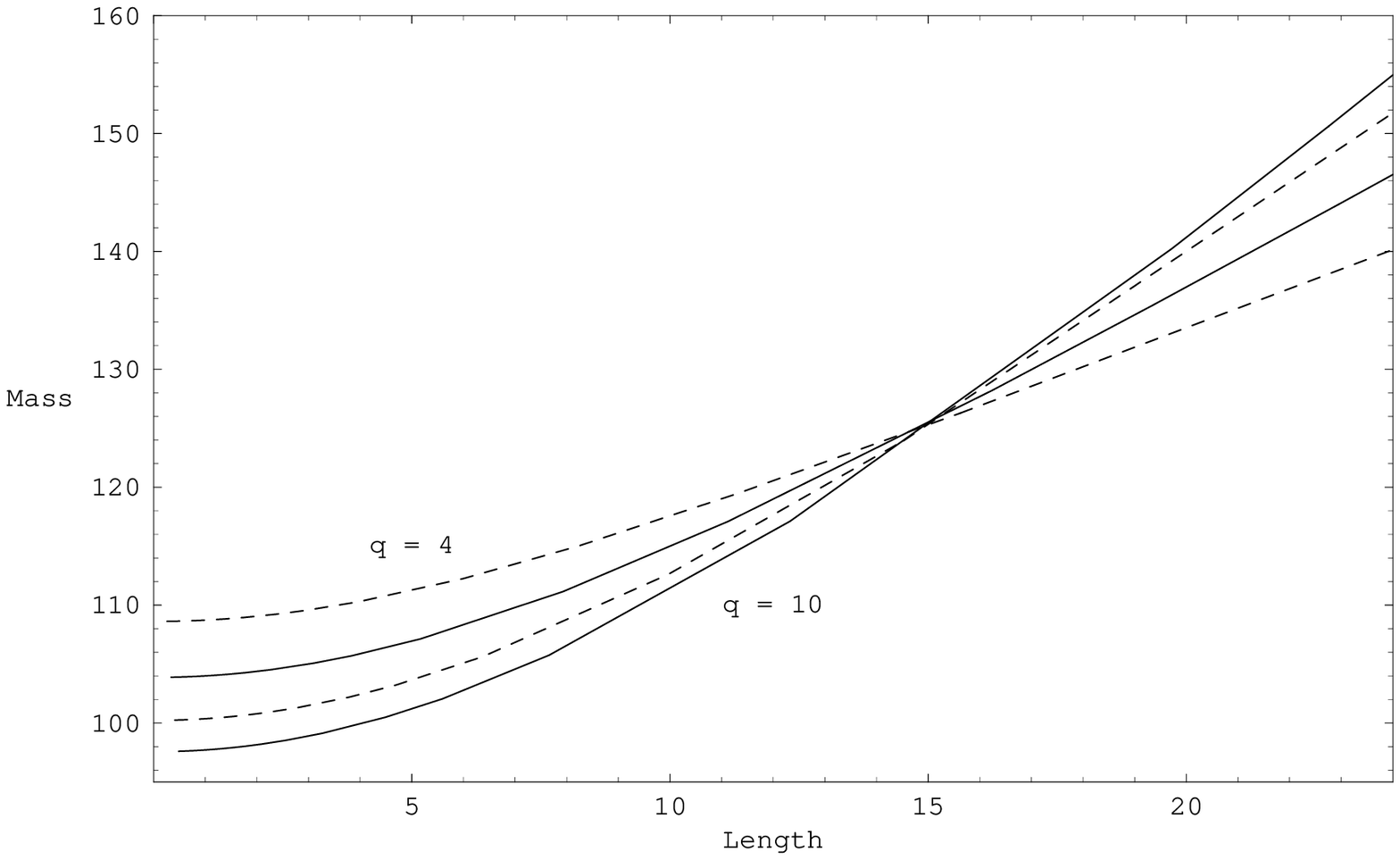,width=8cm}
\end{center}
\noindent {\bf Figure 3:} Mass against length for confining strings
with $q=4,6,8,10$. The background has $N=30$ and $\Phi_0=1$.
\end{figure}

As the length goes to infinity, the tension becomes the known result for infinite
confining strings (\ref{eq:conften}). It is curious that the curves
almost intersect at a single point, call it $(L_0,M_0)$, although
they do not seem to go through precisely the same
point. It seems plausible that the intersection is exact to subleading
order at large $\Delta x$. Mathematically, this fact requires that the subleading
correction to (\ref{eq:conf}) take the following form
\be\label{eq:subleading}
M(q) \sim \frac{e^{\Phi_0} N}{2\pi^2 \alpha'} \sin \frac{\pi q
}{N} \, \left[ \Delta x -  L_0 \right] + M_0 + \cdots  , \qquad \text{as}\qquad \Delta x \to
\infty \,.
\ee
This is an interesting result that completely determines the $q$ dependence of
the subleading term. It seems
difficult to derive (\ref{eq:subleading}) directly from the integral for the mass.

Note that in the other limit, $\Delta x \to 0$, there is no mass divergence in
these cases. This limit is $C \to \infty$ at fixed positive $q$.

It is useful to write the expression for mass (\ref{eq:tens}) in terms of $q$
\be\label{eq:massagain}
M = \frac{N^{3/2} e^{\Phi_0}}{2\pi^2 \a'^{1/2}} \int_0^{\pi}
\frac{\sin^2\psi-(\psi-\pi q /N)(2\sin\psi\cos\psi-\psi+\pi q/N)}{
\sqrt{-e^{2\Phi_0}/C^4 + \sin^4\psi + (\sin\psi \cos\psi -
    \psi + \pi q/N)^2 }} d\psi
\ee
It is simple to check that this integral is invariant under $q
\leftrightarrow N - q$, as we should expect. If $q$ is positive, then
the allowed range of $C^2$ is $[e^{\Phi_0}/\sin(\pi q/N),\infty)$.

Figure 3 and equation (\ref{eq:massagain}) constitute concrete
predictions for the mass of confining strings with finite length. The
subleading, constant, contribution to the mass in (\ref{eq:subleading})
\be
M_0 - \frac{e^{\Phi_0} N}{2\pi^2 \alpha'} \sin \frac{\pi q
}{N} \, L_0 \,,
\ee
is presumably concentrated near the endpoints of the confining string.

\section{IIA background: $G_2$ manifolds}

\subsection{The infrared background: ${\mathcal{M}}^4\times S^2$}

The ${\mathbb{D}}_7$ family of $G_2$ holonomy manifolds are classical
solutions of M
theory
\cite{Cvetic:2001ih,Cvetic:2001kp,Brandhuber:2001kq,Edelstein:2002zy}.
From a IIA perspective, they describe the result of the geometric
transition induced by
D6-branes wrapping an $S^3$ in the deformed conifold. Like the
Maldacena-Nu\~nez background, the solution preserves 4 supercharges
and is thought to be dual to ${\mathcal{N}}=1$ super Yang Mills theory, modulo
issues of decoupling of Kaluza-Klein and gravitational modes
\cite{Gursoy:2003hf}.

In the infrared regime, $r\rightarrow
0$, the background collapses to ${\mathcal{M}}^4\times S^2$ with
$N$ units of RR flux through the sphere\footnote{There is a factor of
  2 missing in the IIA expressions of \cite{Brandhuber:2001kq}. The angle
  of the M theory circle should be rescaled to have range $2\pi$.}
\bea
\label{2afields}
ds_{\text{IIA}}^2 & = & e^{2\Phi_0} \left( dx_{1,3}^2 + \alpha' N^2 \ft{1}{4}
\left[ d\theta^2  + \sin^2 \theta \,d \phi^2 \right] \right) \nn
C_1^{RR} & = & \alpha'^{1/2} N\ft{1}{2}\cos\theta d \phi \,.
\eea
The RR two-form flux is
\be
G_2^{RR} = dC_1^{RR} = -\alpha'^{1/2} N \ft{1}{2} \sin \theta \,
d\theta\wedge d\phi = -\alpha'^{1/2} N \ft{1}{2}
\text{vol}_{S^2} \,.
\ee
The ranges of the angles are $0 \leq \q \leq \pi$ and $0 \leq \p <
2\pi$. The general ${\mathbb{D}}_7$ solution has another free
parameter at $r=0$ which determines the squashing of an $S^3$ in
the $G_2$ geometry. We have set this parameter to 1 for
simplicity. Further, we have rescaled the Minkowski coordinates to
emphasize the similarity with the infrared of the
Maldacena-Nu\~nez solution (\ref{smallr}).

The action for a probe D2-brane with these fluxes is
\be\label{eq:d2action}
S_{\text{DBI}} = - T_2 \int d^3\xi e^{-\Phi} \, \sqrt{ -\det
(^\star G +
  {\mathcal F} ) } + T_2 \int {\mathcal F} \wedge ^\star C_1^{RR} \,.
\ee
As before, setting $r(\xi)=0$ is a consistent truncation of the
full probe brane equations of motion. Therefore we may use the DBI
action in the background at $r=0$ given in (\ref{2afields}).

\subsection{Probe D2-brane solutions}

We will find probe D2-brane solutions similar to the D3-brane
solutions of the previous section. The ansatz we take
describes fundamental strings extended in the $x$ direction blown up
to a D2-brane by the Emparan-Myers effect
\cite{Emparan:1997rt,Myers:1999ps}. At each value of $x$
the D2-brane is `wrapping' an $S^1$ in the $S^2$ of the background.
\bea
t & = & \a'^{1/2} \xi_0 \,, \qquad x = \a'^{1/2} \xi_1 \,, \qquad
\q = \q(\xi_1) \,, \qquad \p =
\xi_2 \,, \nonumber \\
A & = & k(\xi_1) \xi_0 d\xi_1 \qquad \Rightarrow \qquad F =
k(\xi_1) d\xi_0 \wedge d\xi_1 \,.
\eea
One finds that the DBI equations of motion are solved if the
functions $k(\xi_1)$ and $\q(\xi_1)$ satisfy the following
relations
\bea\label{g2soln}
\cos^2\q & = & \kk^2 \,, \nonumber \\
\left[\frac{d\kk}{d\xi_1}\right]^2 & = & C^2 \left[D+\kk\right]
\left[1-\kk^2\right] = C^2 \left[D+\kk\right] \sin^2\q \,,
\eea
where we have introduced
\be
\kk = \frac{32\pi^2 k_0}{N^2 e^{4\Phi_0} C^2} (k-k_0) \,,
\ee
and the three dimensionless constants $C,D,k_0$ are related by
\be
D = \frac{4}{N^2e^{4\Phi_0}C^2} \left[\frac{N^4 e^{8\Phi_0}
    C^4}{256\pi^2 k_0^2} - e^{4\Phi_0} + 4\pi^2 k_0^2 \right] \,.
\ee
The solution therefore has two arbitrary constants. The range of $D$
is restricted to $[1,\infty)$.

We have found a two parameter family of solutions to the full second
order equations of motion. It seems very likely that these solutions
are not supersymmetric, as we showed for the IIB solutions
in the Appendix. However, we have not explicitly checked
non-supersymmetry in the present IIA case.

From the equations (\ref{g2soln}), we see that there will be
solutions where $\q$ runs from $0$ to $\pi$. This corresponds to
$\kk$ running between $-1$ and $1$. The spatial sections of these
solutions are topologically $S^2$, with the worldvolume $S^2$ in
the nontrivial homology class of the background
$H_2({\mathcal{M}}\times S^2)$.

The length of the solutions in the $x$ direction is given by an
elliptic integral
\be
\Delta\, x = \frac{\a'^{1/2}}{C} \int_{-1}^1
\frac{d\kk}{\sqrt{(1-\kk^2)(\kk+D)}} \,.
\ee
It is clear that the length is inversely
proportional to $C$. We may use asymptotic properties of elliptic
integrals to calculate the length as $D\to 1$ and as $D\to\infty$
\bea
\Delta\, x & \sim & \frac{\a'^{1/2}\sqrt{2}}{2C} \ln (D-1) + \cdots \,, \qquad
\text{as} \qquad D \to 1 \,, \nonumber \\
\Delta\, x & \sim & \frac{\a'^{1/2}\pi}{C} \frac{1}{D^{1/2}} + \cdots \,, \qquad
\text{as} \qquad D  \to \infty \,,
\eea
corresponding to long and short extensions respectively.
In the long solutions, only the $\kk = -1$ end goes to infinity, the
$\kk = 1$ point remains at a finite position.

The induced spatial metric of the solutions is given by
\be
ds^2_{\text{D2}} = \frac{\a' N^2}{4} e^{2\Phi_0} \left(
\left[\frac{4}{N^2}+C^2(D+\kk(\xi_1))  \right] d\xi_1^2
+ \left[1-\kk(\xi_1)^2 \right] d\xi_2^2 \right) \,.
\ee
As $\kk\to \pm 1$ one can see that the induced geometry has a conical
singularity due to an angular deficit. As previously, we
interpret the singularities as due to the presence of fundamental
string sources. In this case, the electric charges at the conical points are
\bea
Q^0_{0} & = & \lim_{\q \to 0}
\int_{\xi_2=0}^{2\pi} \frac{\pa
  {\mathcal{L}}}{\pa \pa_1 A_{0}} d\xi_2  \,, \nonumber \\
Q^0_{\pi} & = & - \lim_{\q \to \pi}
\int_{\xi_2=0}^{2\pi} \frac{\pa
  {\mathcal{L}}}{\pa \pa_1 A_{0}} d\xi_2 \,.
\eea
Applying this formula to our solutions, we find
\be
Q^0_0 = \frac{N}{2} - \frac{16 \pi^2 k_0^2}{N C^2 e^{4\Phi_0}} \,,
\qquad \qquad Q^0_{\pi} = \frac{N}{2} + \frac{16 \pi^2 k_0^2}{N C^2 e^{4\Phi_0}} \,.
\ee
The total charge of the sources is $N$, which again precisely
cancels the contribution of the Wess-Zumino term
as required. It will be useful to write $Q^0_{\pi} = N/2
+ q$ and $Q^0_0 = N/2 - q$ with $N/2 \pm q
\in \bZ$. The total number of fundamental strings ending on
the probe brane is again $N$, so the solutions may be interpreted as
deformed baryon vertices.

The four dimensional tension of the solution is given by
\be
T(\xi_1) = \int d\xi_2 {\mathcal{E}} = \frac{4 |k_0|}{\a'^{1/2} N}
\left[\frac{1}{C^2} + \frac{N^2 D}{4} + \frac{N^2 \kk}{4} \right] \,.
\ee
Integrating the tension over the spatial worldvolume gives the mass
\be
M = \int d\xi_1 T(\xi_1) = \frac{4 |k_0|}{\a' N C^2}
\Delta x + \frac{|k_0| N}{\a'^{1/2} C} \int_{-1}^1
d\kk \sqrt{\frac{D+\kk}{(1-\kk^2)}} \,.
\ee
Considering the short and long limits, $D \to \infty$ and $D \to
1$ respectively, one obtains relations between the mass and the
length. There are in fact two possible short limits
\bea
M & \sim & \frac{\a'^{1/2} N^2 e^{2\Phi_0} \pi^2}{4 C^2} \frac{1}{(\Delta x)^2} +
\cdots\,, \qquad \text{as} \qquad \Delta x \to 0 \quad [k_0 \to
\infty \Rightarrow D \to \infty] \,, \nonumber \\
M & \sim & \frac{N^2 e^{2\Phi_0}}{8 \a'^{1/2}} + \cdots\,,
\qquad\qquad
\text{as} \qquad \Delta x \to 0 \quad [k_0 \to
0 \Rightarrow D \to \infty] \,.
\eea
As in the previously discussed IIB solutions, both of the short limits take
us outside the regime for validity of the DBI action.
The long limit is also precisely as for the IIB solutions.
One may obtain the following relation
\be\label{eq:IIAlong}
M \sim \frac{e^{2\Phi_0}}{2\pi \a'} \left|\frac{N}{2}-q\right| \Delta x + \cdots \,,
\qquad \text{as} \qquad \Delta x \to \infty \quad [D \to 1] \,,
\ee
which is exactly the mass of $N/2-q$ fundamental strings in the background
(\ref{2afields}). Again this is consistent with the fact that most of
the mass comes from the $\q=0$ region, where the D2-brane has
collapsed to $N/2-q$ fundamental strings.

Unlike in the Maldacena-Nu\~nez case, the large length limit which
collapses at one end (\ref{eq:IIAlong}) is the only possible large
length limit. This follows from the fact that in (\ref{g2soln}) we
see that $d\xi_1/d\kk$ can only have poles at the endpoints $\kk =
\pm 1$ and not in the interior. Therefore, there is no limit of
the solution space analogous to that of the confining strings we
found previously.

\section{Discussion and conclusions}

We have found explicit nonsupersymmetric probe D-brane solutions in
the infrared of two ${\mathcal{N}}=1$ confining geometries. There is a two
parameter family of solutions which may be labelled by an extension in
spacetime, $\Delta x$, and by a fraction of quarks, $q/N$, that is
being pulled apart from the others in spacetime.

The solutions describe deformed baryon vertices. In the IIB case we
considered, the Maldacena-Nu\~nez background, there was a limit at
large $\Delta x$ in which the solutions became the infinite
confining strings of the dual theory. Away from the infinite length
limit, the solutions give predictions for the mass of finite confining
strings in the dual theory.

We found a similar two parameter family of solutions in a IIA
geometry, obtained by dimensional reduction of a $G_2$ holonomy
background of M theory. However,
in the IIA solutions there is not a limit
with the properties of infinite confining strings.
The only large length limits involve collapse into fundamental strings
at one end of the baryon vertex.
There does exist a proposal for identifying the confining strings as membranes
in the $G_2$ background \cite{Acharya:2001hq}. Translating this idea
into a formula for the string tensions with the required symmetry $q
\leftrightarrow N - q$ remains an open problem.

Various possibilities for future research suggest themselves.
It seems likely that solutions similar to those we have described will
exist in other backgrounds, including $AdS_5 \times S^5$. In fact,
presumably a systematic study of such DBI solutions in infrared
geometries of the form ${\mathcal{M}}^4\times S^{8-p}$ is possible,
along the lines of \cite{Camino:1999xx}. The fact that the infrared
geometry is independent of how the compact directions of the
background D-brane are wrapped highlights the genericity of the
solutions and of the dual field theory deformed baryons/confining
strings.

It would be interesting to see if other appearances of confining
strings in dualities admit similar energetics at finite length.
Important examples are the confining strings of MQCD
\cite{Hanany:1997hr,Douglas:1995nw} and of the theory on nonextremal D4
branes at high temperature \cite{Callan:1999zf}.

Ultimately, one would like to reproduce the properties of finite
confining strings that we have described via a field theoretic
calculation. Recent field theory work on confining strings in four
dimensional $SU(N)$ theories has been both numerical, see for example
\cite{Lucini:2004my}, and analytical \cite{Shoshi:2002rd}.

\noindent {\bf Acknowledgements}:  The authors would like to thank
David Berman, Jos\'e Edelstein, Prem Kumar, Carlos N\'{u}\~{n}ez,
Alfonso Ramallo, Konstantin
Savvidy, James Sparks and Steffan Theissen for interesting and
helpful conversations. S.A.H. is funded by the Sims
scholarship. R.P. would like to acknowledge
the generous support to Centro de Estudios Cientificos (CECS) by Empresas
CMPC. CECS is a Millenium Science Institute and is
funded in part by grants from Fundaci\'on Andes and the Tinker Foundation.

\appendix

\section{Checking non-supersymmetry of IIB solutions}

A probe brane is supersymmetric if at least one of the Killing
spinors of the background, $\e$, satisfies
\be\label{kappa2}
\G_\kappa \e = \e \,,
\ee
where \cite{Bergshoeff:1996tu}
\be
\Gamma_\kappa =
\frac{i}{\sqrt{^\star G+F}}\sum_0^\infty \frac{1}{2^n n!} \gamma^{\mu_1
\nu_1 \dots \mu_n \nu_n} F_{\mu_1 \nu_1}\dots F_{\mu_n \nu_n}
J^{(n)}_p\,.
\ee
In type IIB supergravity we have
\be
J^{(n)}_p = (-1)^n \left[ (\sigma_3)^{n + \frac{p-3}{2}}
\sigma_2 \otimes \Gamma_{(0)}\right]  \,,
\ee
with
\be
\Gamma_{(0)} = \frac{1}{(p+1)!} \epsilon^{\mu_1 \dots
\mu_{(p+1)}} \gamma_{\mu_1 \dots \mu_{(p+1)}} \,.
\ee
We are using the standard notation in which we write the IIB
spinors as an $SL(2,\bR)$ doublet of real spinors. The lower case
gamma matrices are the pull-back of the spacetime gamma matrices
$\g_\mu = E^{\bar{a}}{}_{\mu} \G_{\bar{a}}$.

If we use the following vielbein for the metric at the origin
(\ref{smallr})
\bea
E^i & = & e^{\Phi_0/2} dx^i \qquad (i=0,1,2,3) \,, \nonumber \\
E^4 & = & 0 \,, \nonumber \\
E^5 & = & 0 \,, \nonumber \\
E^6 & = & 0 \,, \nonumber \\
E^7 & = & e^{\Phi_0/2} d\psi \,, \nonumber \\
E^8 & = & e^{\Phi_0/2} \sin\psi d\theta \,, \nonumber \\
E^9 & = & e^{\Phi_0/2} \sin\psi \sin\theta d\phi \,,
\eea
then the supersymmetry projector for the embedding
(\ref{eq:d3embedding}) becomes
\be\label{kappa1}
\Gamma_\kappa = \frac{i}{\sqrt{k^2
- e^{2\Phi_0}(1+(\pa_1\psi)^2)}}\left[ e^{\Phi_0}(\sigma_2 \otimes
\G_0[\G_1 + \pa_1 \psi \G_7]\G_{89} ) + k (\sigma_3 \sigma_2
\otimes \G_{89}) \right]\,.
\ee

The Killing spinors of the Maldacena-Nu\~{n}ez background are
given in \cite{Nunez:2003cf}. There are four real supercharges.
However, the only property we shall need is that the spinors
satisfy
\be \label{back}
\left(\sigma_1 \otimes \bbbone\right) \e = \e \,.
\ee
If we use this property in (\ref{kappa2}) we find that a necessary
condition for solutions is that $\left(\G_1 + \pa_1\psi
\G_7\right) \e = 0$. However this condition then requires $\pa_1 \psi = \pm
1$, which is not consistent with the form of the solutions
(\ref{MNsolutions}). Therefore, none of the solutions are
supersymmetric.


\begin{thebibliography}{99}

\bibitem{Maldacena:1997re}
J.~M.~Maldacena, ``The large N limit of superconformal field
theories and supergravity,'' Adv.\ Theor.\ Math.\ Phys.\  {\bf 2}
(1998) 231 [Int.\ J.\ Theor.\ Phys.\  {\bf 38} (1999) 1113]
[arXiv:hep-th/9711200].

\bibitem{Gubser:1998bc}
S.~S.~Gubser, I.~R.~Klebanov and A.~M.~Polyakov, ``Gauge theory
correlators from non-critical string theory,'' Phys.\ Lett.\ B
{\bf 428} (1998) 105 [arXiv:hep-th/9802109].

\bibitem{Witten:1998qj}
E.~Witten, ``Anti-de Sitter space and holography,'' Adv.\ Theor.\
Math.\ Phys.\  {\bf 2} (1998) 253 [arXiv:hep-th/9802150].

\bibitem{Klebanov:2000hb}
I.~R.~Klebanov and M.~J.~Strassler,
 ``Supergravity and a confining gauge theory: Duality cascades and
JHEP {\bf 0008} (2000) 052 [arXiv:hep-th/0007191].

\bibitem{Maldacena:2000yy}
J.~M.~Maldacena and C.~Nunez, ``Towards the large N limit of pure
N = 1 super Yang Mills,'' Phys.\ Rev.\ Lett.\  {\bf 86} (2001) 588
[arXiv:hep-th/0008001].

\bibitem{Witten:1998xy}
E.~Witten, ``Baryons and branes in anti de Sitter space,'' JHEP
{\bf 9807} (1998) 006 [arXiv:hep-th/9805112].

\bibitem{Gross:1998gk}
D.~J.~Gross and H.~Ooguri, ``Aspects of large N gauge theory
dynamics as seen by string theory,'' Phys.\ Rev.\ D {\bf 58}
(1998) 106002 [arXiv:hep-th/9805129].

\bibitem{Callan:1998iq}
C.~G.~Callan, A.~Guijosa and K.~G.~Savvidy, ``Baryons and string
creation from the fivebrane worldvolume action,'' Nucl.\ Phys.\ B
{\bf 547} (1999) 127 [arXiv:hep-th/9810092].

\bibitem{Imamura:1998gk}
Y.~Imamura, ``Supersymmetries and BPS configurations on Anti-de
Sitter space,'' Nucl.\ Phys.\ B {\bf 537} (1999) 184
[arXiv:hep-th/9807179].

\bibitem{Gomis:1999xs}
J.~Gomis, A.~V.~Ramallo, J.~Simon and P.~K.~Townsend,
``Supersymmetric baryonic branes,'' JHEP {\bf 9911} (1999) 019
[arXiv:hep-th/9907022].

\bibitem{Callan:1999zf}
C.~G.~Callan, A.~Guijosa, K.~G.~Savvidy and O.~Tafjord, ``Baryons
and flux tubes in confining gauge theories from brane actions,''
Nucl.\ Phys.\ B {\bf 555} (1999) 183 [arXiv:hep-th/9902197].

\bibitem{Herzog:2001fq}
C.~P.~Herzog and I.~R.~Klebanov,
``On string tensions in supersymmetric SU(M) gauge theory,''
Phys.\ Lett.\ B {\bf 526} (2002) 388
[arXiv:hep-th/0111078].

\bibitem{Pawelczyk:2000ah}
J.~Pawelczyk, ``SU(2) WZW D-branes and their noncommutative
geometry from DBI action,'' JHEP {\bf 0008} (2000) 006
[arXiv:hep-th/0003057].

\bibitem{Pawelczyk:2000hy}
J.~Pawelczyk and S.~J.~Rey, ``Ramond-Ramond flux stabilization of
D-branes,'' Phys.\ Lett.\ B {\bf 493} (2000) 395
[arXiv:hep-th/0007154].

\bibitem{Bachas:2000ik}
C.~Bachas, M.~R.~Douglas and C.~Schweigert,
``Flux stabilization of D-branes,''
JHEP {\bf 0005} (2000) 048
[arXiv:hep-th/0003037].

\bibitem{Cvetic:2001ih}
M.~Cvetic, G.~W.~Gibbons, H.~Lu and C.~N.~Pope,
``M-theory conifolds,''
Phys.\ Rev.\ Lett.\  {\bf 88} (2002) 121602
[arXiv:hep-th/0112098].

\bibitem{Cvetic:2001kp}
M.~Cvetic, G.~W.~Gibbons, H.~Lu and C.~N.~Pope, ``A G(2)
unification of the deformed and resolved conifolds,'' Phys.\
Lett.\ B {\bf 534} (2002) 172 [arXiv:hep-th/0112138].

\bibitem{Brandhuber:2001kq}
A.~Brandhuber,
``G(2) holonomy spaces from invariant three-forms,''
Nucl.\ Phys.\ B {\bf 629} (2002) 393
[arXiv:hep-th/0112113].

\bibitem{Edelstein:2002zy}
J.~D.~Edelstein, A.~Paredes and A.~V.~Ramallo,
``Let's twist again: General metrics of G(2) holonomy from gauged
supergravity,''
JHEP {\bf 0301} (2003) 011
[arXiv:hep-th/0211203].

\bibitem{Gursoy:2003hf}
U.~Gursoy, S.~A.~Hartnoll and R.~Portugues,
``The chiral anomaly from M theory,''
Phys.\ Rev.\ D {\bf 69} (2004) 086003
[arXiv:hep-th/0311088].

\bibitem{Loewy:2001pq}
A.~Loewy and J.~Sonnenschein, ``On the holographic duals of N = 1
gauge dynamics,'' JHEP {\bf 0108} (2001) 007
[arXiv:hep-th/0103163].

\bibitem{Acharya:2001dz}
B.~S.~Acharya and C.~Vafa, ``On domain walls of N = 1
supersymmetric Yang-Mills in four dimensions,''
arXiv:hep-th/0103011.

\bibitem{Emparan:1997rt}
R.~Emparan,
``Born-Infeld strings tunneling to D-branes,''
Phys.\ Lett.\ B {\bf 423} (1998) 71
[arXiv:hep-th/9711106].

\bibitem{Myers:1999ps}
R.~C.~Myers,
``Dielectric-branes,''
JHEP {\bf 9912} (1999) 022
[arXiv:hep-th/9910053].

\bibitem{Camino:1999xx}
J.~M.~Camino, A.~V.~Ramallo and J.~M.~Sanchez de Santos,
``Worldvolume dynamics of D-branes in a D-brane background,''
Nucl.\ Phys.\ B {\bf 562} (1999) 103
[arXiv:hep-th/9905118].

\bibitem{Camino:2001at}
J.~M.~Camino, A.~Paredes and A.~V.~Ramallo,
``Stable wrapped branes,''
JHEP {\bf 0105} (2001) 011
[arXiv:hep-th/0104082].

\bibitem{Callan:1997kz}
C.~G.~Callan and J.~M.~Maldacena,
``Brane dynamics from the Born-Infeld action,''
Nucl.\ Phys.\ B {\bf 513} (1998) 198
[arXiv:hep-th/9708147].

\bibitem{Gibbons:1997xz}
G.~W.~Gibbons,
``Born-Infeld particles and Dirichlet p-branes,''
Nucl.\ Phys.\ B {\bf 514} (1998) 603
[arXiv:hep-th/9709027].

\bibitem{Hanany:1997hr}
A.~Hanany, M.~J.~Strassler and A.~Zaffaroni,
``Confinement and strings in M{QCD},''
Nucl.\ Phys.\ B {\bf 513} (1998) 87
[arXiv:hep-th/9707244].

\bibitem{Douglas:1995nw}
M.~R.~Douglas and S.~H.~Shenker,
``Dynamics of SU(N) supersymmetric gauge theory,''
Nucl.\ Phys.\ B {\bf 447} (1995) 271
[arXiv:hep-th/9503163].

\bibitem{Acharya:2001hq}
B.~S.~Acharya,
``Confining strings from G(2)-holonomy spacetimes,''
arXiv:hep-th/0101206.

\bibitem{Lucini:2004my}
B.~Lucini, M.~Teper and U.~Wenger,
``Glueballs and k-strings in SU(N) gauge theories: Calculations with improved
operators,''
arXiv:hep-lat/0404008.

\bibitem{Shoshi:2002rd}
A.~I.~Shoshi, F.~D.~Steffen, H.~G.~Dosch and H.~J.~Pirner,
``Confining QCD strings, Casimir scaling, and a Euclidean approach to
high-energy scattering,''
Phys.\ Rev.\ D {\bf 68} (2003) 074004
[arXiv:hep-ph/0211287].

\bibitem{Bergshoeff:1996tu}
E.~Bergshoeff and P.~K.~Townsend,
``Super D-branes,''
Nucl.\ Phys.\ B {\bf 490} (1997) 145
[arXiv:hep-th/9611173].

\bibitem{Nunez:2003cf}
C.~Nunez, A.~Paredes and A.~V.~Ramallo,
``Flavoring the gravity dual of N = 1 Yang-Mills with probes,''
JHEP {\bf 0312} (2003) 024
[arXiv:hep-th/0311201].

\end{thebibliography}
\end{document}